\documentclass[conference]{IEEEtran}
\IEEEoverridecommandlockouts
\usepackage{cite}
\usepackage{amsmath,amssymb,amsfonts}
\usepackage{algorithmic}
\usepackage{graphicx}
\usepackage{textcomp}
\usepackage{xcolor}
\usepackage{threeparttable}
\usepackage{multirow}
\def\BibTeX{{\rm B\kern-.05em{\sc i\kern-.025em b}\kern-.08em
    T\kern-.1667em\lower.7ex\hbox{E}\kern-.125emX}}

\makeatletter
\newcommand{\linebreakand}{
\end{@IEEEauthorhalign}
\hfill\mbox{}\par
\mbox{}\hfill\begin{@IEEEauthorhalign}
}
\makeatother

\begin{document}

\title{TLSQKT: A Question-Aware Dual-Channel Transformer for Literacy Tracing from Learning Sequences\\

\thanks{Corresponding author: Chunyan Zeng, Email: cyzeng@hbut.edu.cn.}
}

\author{\IEEEauthorblockN{Zhifeng Wang}
	\IEEEauthorblockA{\textit{Faculty of Artificial Intelligence in Education} \\
		\textit{Central China Normal University}\\
		Wuhan 430079, China \\
		zfwang@ccnu.edu.cn}
	\and
	\IEEEauthorblockN{Yaowei Dong}
	\IEEEauthorblockA{\textit{Faculty of Artificial Intelligence in Education} \\
		\textit{Central China Normal University}\\
		Wuhan, China \\
		kinglove@mails.ccnu.edu.cn}
	\linebreakand
	\IEEEauthorblockN{Chunyan Zeng}
	\IEEEauthorblockA{\textit{School of Electrical and Electronic Engineering} \\
		\textit{Hubei University of Technology}\\
		Wuhan 430068, China \\
		cyzeng@hbut.edu.cn}
}



\maketitle

\IEEEpubidadjcol

\begin{abstract}
Knowledge tracing (KT) supports personalized learning by modeling how students’ knowledge states evolve over time. However, most KT models emphasize mastery of discrete knowledge components, limiting their ability to characterize broader literacy development. We reframe the task as Literacy Tracing (LT), which models the growth of higher-order cognitive abilities and literacy from learners’ interaction sequences, and we instantiate this paradigm with a Transformer-based model, TLSQKT (Transformer for Learning Sequences with Question-Aware Knowledge Tracing). TLSQKT employs a dual-channel design that jointly encodes student responses and item semantics, while question-aware interaction and self-attention capture long-range dependencies in learners’ evolving states. Experiments on three real-world datasets—one public benchmark, one private knowledge-component dataset, and one private literacy dataset—show that TLSQKT consistently outperforms strong KT baselines on literacy-oriented metrics and reveals interpretable developmental trajectories of learners’ literacy. Transfer experiments further indicate that knowledge-tracing signals can be leveraged for literacy tracing, offering a practical route when dedicated literacy labels are limited. These findings position literacy tracing as a scalable component of intelligent educational systems and lay the groundwork for literacy evaluation in future large-scale educational models.
\end{abstract}


\section{Introduction}
Personalized learning in intelligent educational systems depends on accurately modeling how students’ knowledge evolves as they interact with learning materials \cite{Dong2023}. Knowledge tracing addresses this need by inferring latent knowledge states from historical learning records and using these states to predict performance on future items \cite{Shi2026,Wang2025b,Dong2025,Chen2025b,Liao2024,Ma2023b}. Classical approaches such as Bayesian Knowledge Tracing (BKT) represent knowledge as probabilistic mastery of predefined skills \cite{BKT}. With the advent of deep learning, models such as Deep Knowledge Tracing (DKT) have shifted KT toward neural sequence modeling, which enables richer representations of student–item interactions and improved predictive accuracy \cite{DKT}. This progression supports large-scale, automated assessment pipelines that reduce teachers’ routine workload and deliver timely, personalized feedback to learners \cite{Chen2024f,Wang2025d,Li2023g,Wang2024r,Chen2024e,Wang2024q}.

Despite these advances, the predominant KT paradigm remains focused on mastery of discrete knowledge components (KCs) \cite{Li2026a,Lyu2022,Wang2023h,Li2023j,Wang2023c,Li2023e}. This focus limits the ability to characterize broader, higher-order competencies that describe students’ literacy and cognitive development. In many curricula, core literacy comprises multiple interrelated dimensions \cite{Liao2024}. For example, high school mathematics literacy is often described by six dimensions: mathematical abstraction, logical reasoning, mathematical modeling, spatial imagination, mathematical computation, and data analysis. Modeling only item–skill mastery obscures cross-skill transfer, metacognitive strategies, and other holistic abilities that contribute to long-term learning outcomes \cite{Li2023j}. Evaluation practices that emphasize short-horizon next-item prediction also provide limited insight into learners’ developmental trajectories and the interpretability required for formative assessment.

These observations motivate a shift from traditional knowledge tracing to \textit{Literacy Tracing} (LT). Literacy tracing aims to model the development of higher-order cognitive abilities and literacy by leveraging learners’ interaction sequences in a manner that complements KC-level mastery. The goal is to estimate broader literacy states, reveal their trajectories over time, and produce diagnostics that are actionable for instruction and policy. This reframing retains the predictive strengths of KT while aligning model outputs with constructs that educators and curriculum standards prioritize.

Advancing from KT to LT introduces several technical and practical challenges. First, there is multidimensionality and dependency structure. Literacy dimensions interact in complex ways across items, topics, and time, which requires models that can capture long-range dependencies and cross-dimension correlations without collapsing them into a single index. Second, there is label scarcity and construct validity. Literacy labels are costly to obtain and must align with well-defined constructs. Models should therefore be designed to transfer knowledge learned from KT tasks to LT tasks and to remain robust when explicit literacy supervision is limited. Third, there is temporal granularity and interpretability. Literacy grows cumulatively and nonlinearly. Systems must produce temporally coherent estimates and offer interpretable evidence that can inform instructional decisions rather than only yield opaque predictions.

To address these challenges, we propose a knowledge–literacy collaborative diagnostic framework that simultaneously assesses students at the KC level and at the literacy level. Within this framework, we instantiate a Transformer-based model, TLSQKT (Transformer for Learning Sequences with Question-Aware Knowledge Tracing). TLSQKT integrates the sequential modeling capacity of the Transformer architecture \cite{attention} with a dual-channel strategy inspired by QIKT \cite{QIKT}, enabling the model to couple item semantics with learner responses. Question-aware interaction strengthens the alignment between item intent and the evolving learner state, while self-attention captures long-range dependencies that govern how earlier experiences shape subsequent literacy development. The design is compatible with existing KT pipelines but extends them to literacy-oriented targets and analyses, which supports unified deployment in real educational systems.

From a methodological perspective, our approach contributes three elements that together support scalable LT. First, we formalize literacy tracing as a complementary objective to knowledge tracing and clarify how to construct literacy-oriented targets from learning sequences and auxiliary annotations. Second, we provide a model architecture that can be trained on KT data and adapted to LT with minimal overhead, which directly addresses label scarcity and facilitates cross-task transfer. Third, we outline evaluation protocols that move beyond next-item accuracy to include literacy-oriented metrics and trajectory analyses that assess calibration, stability, and interpretability.

\begin{enumerate}
	\item \textbf{Framework.} We introduce a collaborative diagnostic framework that aligns KC-level mastery with multidimensional literacy states, offering comprehensive learner profiles suitable for formative assessment and personalization.
	\item \textbf{Model.} We design TLSQKT, a Transformer-based sequence model with dual-channel encoding and question-aware interaction that jointly leverages item semantics and learner responses for both KT and LT.
	\item \textbf{Empirical validation.} We evaluate our approach on three real-world datasets that include a public benchmark, a private KC dataset, and a private literacy dataset. Results show improved performance over strong KT baselines on literacy-oriented metrics and reveal interpretable developmental trajectories of learners’ literacy.
\end{enumerate}

The remainder of the paper is organized as follows. Section~\ref{section2} reviews related work on knowledge tracing, cognitive diagnosis, and subject-specific analyses of core literacy. Section~\ref{section3} defines the KT and LT problems and presents the proposed model and its components. Section~\ref{section4} details the experimental setup and results, followed by a discussion of findings and implications. Section~\ref{section5} concludes and outlines directions for future work.

\section{Related Work}\label{section2}

This section reviews three strands that ground literacy tracing: knowledge tracing , cognitive diagnosis, and disciplinary core literacy. 

\subsection{Knowledge Tracing}

Knowledge tracing aims to infer and update a learner’s latent mastery of knowledge components from sequential interaction data and to predict future performance. The transition from probabilistic to neural sequence modeling has markedly advanced the field. Deep Knowledge Tracing introduced recurrent architectures to KT, demonstrating that RNNs and LSTMs can capture temporal dependencies in student–item interactions and yield strong predictive performance~\cite{DKT,Lyu2022}. To mitigate limitations of the original DKT (for example, overfitting, vanishing gradients, and limited interpretability), subsequent variants incorporated regularization, explicit forgetting mechanisms, and refined objective functions, including DKT+ and DKT-Forget~\cite{DKT+,DKT-forget}.

Beyond these early neural baselines, a diverse ecosystem of KT models has emerged based on deep learning methods~\cite{Zeng2025b,Wang2025e,Zeng2024b,Zeng2023d,Li2023b,Zeng2022b,Zeng2022,Wang2022n,Zeng2021,Wang2021a,Zeng2025}. Broadly, existing work can be grouped into: (i) \emph{sequence-based models} that refine recurrent architectures or adopt temporal convolutions; (ii) \emph{memory-augmented models} that use external memories to store and retrieve concept-level states; (iii) \emph{forgetting-aware models} that explicitly encode decay and spacing effects to capture realistic learning and forgetting; (iv) \emph{graph-based models} that leverage concept graphs or student–item relations to model structural dependencies; (v) \emph{attention- or Transformer-based models} that use self-attention to capture long-range dependencies and item semantics; and (vi) \emph{context-aware models} that incorporate side information such as item text, difficulty, or instructional context. These lines of work have steadily improved next-item prediction and robustness across datasets, while also clarifying theoretical connections between temporal modeling and latent mastery dynamics.

Despite this progress, several challenges remain central to KT: (a) \emph{multidimensional competence} is difficult to capture when models focus narrowly on KCs; (b) \emph{long-horizon coherence} is not guaranteed when objectives emphasize short-range predictions; and (c) \emph{interpretability for formative assessment} remains limited in purely black-box formulations. These gaps motivate extensions that connect KC-level mastery to broader, higher-order competencies, which we address through literacy tracing.

\subsection{Cognitive Diagnosis}

Cognitive diagnosis provides a complementary measurement perspective that estimates learners’ abilities or attribute mastery with psychometric grounding. Early work in this area followed the ability-level paradigm, modeling proficiency as one- or multi-dimensional latent traits, as in Item Response Theory (IRT) and Multidimensional IRT (MIRT)~\cite{IRT,MIRT}. To obtain fine-grained skill profiles, the Q-matrix and the cognitive-attribute paradigm were introduced, leading to diagnostic models such as the Rule Space Model (RSM), DINA, and the General Diagnostic Model (GDM)~\cite{RSM,DINA,GDM}. These models formalize the mapping between items and attributes, yielding interpretable mastery vectors suitable for diagnostic feedback.

With increasing data availability and advances in machine learning, cognitive diagnosis has expanded to include nonparametric and neural approaches. Classical machine learning methods (for example, clustering, support vector machines, and matrix factorization) have been used to infer latent structures from response data~\cite{cluster,SVM,MatrixFactorization}. More recently, deep learning has enabled richer representation learning and flexible inference pipelines \cite{Zheng2025,Zeng2024g,Chen2025,Zeng2024e,Zheng2024,Zeng2024c,Wang2023l,Zeng2024f,Chen2025a,Zeng2024h,Chen2023b,Zeng2024d,Wang2022c,Zeng2024a,Zeng2024,Zeng2023b,Zeng2023a,Zeng2022a,Zeng2021b,Zeng2020a}. Prior work in this direction can be organized into three threads: (i) \emph{mastery-pattern classifiers}, which cast attribute mastery estimation as a supervised classification task~\cite{MPC}; (ii) \emph{cognitive interaction simulators}, which model response processes with deep neural networks, neural architecture search, or graph neural networks to capture complex dependencies among students, items, and attributes~\cite{DNN,NAS}; and (iii) \emph{encoder–decoder frameworks}, which integrate representation learning with predictive decoding to jointly recover interpretable attributes and response behavior.

Relative to KT, psychometric CDMs offer stronger construct interpretability and principled assessment design, whereas modern KT excels at large-scale sequence prediction. Bridging these traditions remains an active area: integrating CDM-style interpretability with KT’s sequential expressiveness can produce diagnostics that are both predictive and pedagogically actionable.

\subsection{Disciplinary-Specific Core Literacy}

Disciplinary-specific core literacy comprises advanced competencies that integrate domain knowledge, cognitive strategies, and values needed for problem solving in authentic contexts. In secondary mathematics, for example, literacy is often described via multiple interrelated dimensions, including mathematical abstraction, logical reasoning, mathematical modeling, spatial imagination, mathematical computation, and data analysis. These literacies are developmental and context-sensitive: they evolve cumulatively through exposure to curricular practices and tasks, and they manifest across items and topics rather than aligning one-to-one with isolated KCs.

Recent educational research emphasizes embedding core literacy into curriculum and instruction through practices such as project-based learning and inquiry-oriented tasks, which situate knowledge use in meaningful contexts. Assessment research, in parallel, has moved toward diversified systems that combine performance-based tasks with formative assessment to capture process and growth, not only outcomes. However, several challenges persist: (i) \emph{multidimensionality} complicates modeling since literacy dimensions are correlated and may interact nonlinearly; (ii) \emph{temporal accumulation} requires longitudinal analyses that go beyond cross-sectional testing; and (iii) \emph{construct validity} demands that operational measures align with theoretically grounded literacy constructs.

These challenges create an opportunity for computational approaches that connect interaction data with literacy constructs. In particular, KT-style sequence modeling can provide fine-grained temporal signals, while cognitive diagnosis offers interpretable constructs and validity arguments. Our work positions \emph{literacy tracing} at this intersection: we leverage sequence models to estimate and track multidimensional literacy states over time, while maintaining compatibility with KC-level diagnostics for comprehensive, actionable assessment.

\section{Proposed Method}
\label{section3}

\begin{figure*}[htbp]
	\centering
	\includegraphics[width=1\linewidth]{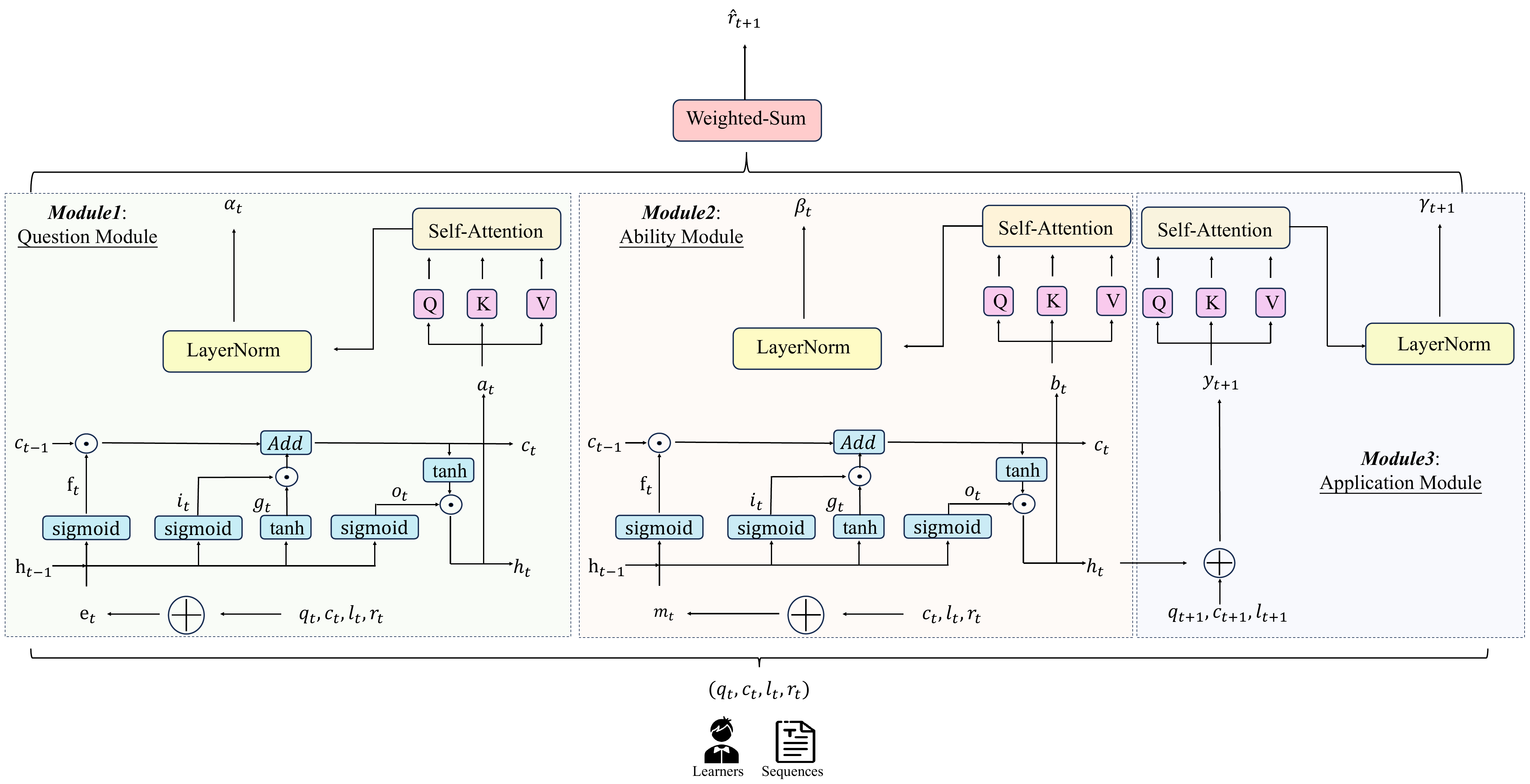}
	\caption{Overview of the proposed Transformer for Learning Sequences with Question-Aware Knowledge Tracing model.}
	\label{fig-model-structure}
\end{figure*}

In this section, we formalize both knowledge tracing and literacy tracing and introduce TLSQKT, a Transformer-based model with question-aware interaction that unifies the two tasks. We first specify the KT setting over sequences of question–concept–response triples and then extend it to LT by associating each interaction with a literacy dimension, yielding quadruples that expose higher-order competencies. Building on this formulation, TLSQKT comprises three coordinated modules—the Question Module, the Ability Module, and the Application Module—that embed interactions from complementary perspectives. Each module uses an LSTM to capture local temporal patterns and a Transformer block to model long-range dependencies and cross-step relations, producing module-specific scores that reflect question-oriented, literacy-oriented, and application-oriented states. A lightweight prediction head linearly combines these scores to estimate next-response correctness, enabling joint exploitation of item semantics, literacy signals, and historical context, as shown in Fig.~\ref{fig-model-structure}. This design preserves compatibility with established KT pipelines while providing a scalable pathway to LT, supporting fine-grained tracking of mastery and growth in multidimensional literacy.

\subsection{Problem Definition}

In the knowledge tracing task, the primary focus is on the model’s performance in predicting students’ responses to questions. A student’s learning activity is typically represented as a sequence of question-response interactions, along with other contextual information. We define a question set $Q$ and a knowledge concept set $C$. For a given student $i$, at time step $t$, the student attempts a question $q_t^i \in Q$, which is associated with a knowledge component $c_t^i \in C$, and provides a response $r_t^i \in \{0,1\}$, indicating whether the student answered the question correctly (1) or incorrectly (0). Therefore, each student has a sequence of interactions denoted as:

\begin{equation}
    \{(q_1,c_1,r_1),...,(q_T,c_T,r_T)\},q_t \in Q,c_t \in C, r_t \in \{0,1\}
\end{equation}

A knowledge tracing model aims to track a student’s knowledge state $h_t$ based on their sequence of learning activities, in order to reflect their current level of knowledge mastery and to predict the student’s response $\hat{r}_{t+1}$ to the next question.

\begin{equation}
    \hat{r}_{t+1}=p(r_{t+1}=1|(q_1,c_1,r_1),...,(q_t,c_t,r_t),(q_t,c_t))
\end{equation}

In the literacy tracing task, we also focus on the model’s performance in predicting students’ responses to questions. However, unlike knowledge tracing, this task emphasizes the literacy dimensions involved in the learning activities. Let L denote the set of competency dimensions. In this setting, each question a student answers is associated not only with a knowledge component $c_t^i \in C$, but also with a corresponding literacy dimension $l_t^i \in L$. Therefore, each student’s literacy interaction sequence consists of a series of four-tuples: $s_t=(q_t,c_t,l_t,r_t).$

We extend the knowledge tracing model to the literacy tracing task by using the literacy interaction sequence to track a student’s literacy state $\lambda_t$, and to predict the student’s response $\hat{r}_{t+1}$ to the next question.

\begin{equation}
    \hat{r}_{t+1}=p(r_{t+1}=1|s_1, s_2,...,s_t,q_{t+1},r_{t+1,},l_{t+1})
\end{equation}

\subsection{Transformer for Learning Sequences with Question-Aware Knowledge Tracing}

To better adapt knowledge tracing models to literacy tracing tasks, we incorporate the strengths of existing KT models. For instance, the QIKT model employs a dual-channel modeling approach that captures learning interactions at both the knowledge-component level and the question level, enabling the model to infer knowledge states from multiple perspectives. Similarly, the DTransformer model introduces a cross-attention mechanism that explicitly models the mastery of knowledge component. Building on these insights, we propose a Transformer-based Knowledge Tracing Model for Learning Sequence with Question-Aware Interaction, as illustrated in Fig. \ref{fig-model-structure}. TLSQKT combines the advantages of dual-channel modeling and long-sequence representation, and demonstrates strong performance on both knowledge tracing and literacy tracing tasks.

\subsubsection{Input Embedding}

Our model consists of three modules: the Question Module, the Ability Module, and the Application Module. These modules process data from different perspectives—question-oriented, ability-oriented, and application-oriented—using distinct embedding strategies. For the Question Module, we consider the interaction from the perspective of the entire question. Therefore, the input is encoded as a combined embedding of the question, literacy dimension, and response:

\begin{equation}
    e_t=Embed(q_t,l_t,r_t)
\end{equation}

For the literacy Module, we approach the interaction from the perspective of the student’s deeper-level abilities. Accordingly, the input is encoded as a combined embedding of the core literacy dimension and the response:

\begin{equation}
    m_t=Embed(l_t,r_t)
\end{equation}

For the Application Module, we consider the interaction from an application-oriented perspective. Its input is influenced not only by the current question and literacy, but also by the student’s historical learning data:

\begin{equation}
    y_{t+1}=Embed(b_t,q_{t+1},l_{t+1})
\end{equation}

\subsubsection{Literacy Tracing}

In the literacy tracing process, we analyze the tracking behavior through each of the three modules individually. For the Question Module, we employ an LSTM to capture the relationships within the learning sequence and obtain a vector representation $a_t$ that reflects the student’s question-related state:

\begin{equation}
    a_t=LSTM(e_t)
\end{equation}

Next, we use a Transformer block to capture the relationships among the question state vectors, and then apply a fully connected layer to obtain the question score $\alpha_t$ for this module:

\begin{equation}
    \alpha_t=out(TransformerLayer(a_t))
\end{equation}

For the ability Module, we similarly use an LSTM to capture the relationships within the student’s learning literacy sequence, resulting in a vector $b_t$ that represents the student’s literacy state:

\begin{equation}
    b_t=LSTM(m_t)
\end{equation}

Next, we also use a Transformer block to capture the relationships among the literacy state vectors, and apply a fully connected layer to obtain the literacy score $\beta_t$ for this module:

\begin{equation}
    \beta_t=out(TransformerLayer(b_t))
\end{equation}

For the Application Module, the input is a combined embedding $y_t$ of the current question, literacy dimension, and the literacy state vector. A Transformer block is used to capture the relationship between the current question and the historical interactions, followed by a fully connected layer to produce the application score $\gamma_{t+1}$:

\begin{equation}
    \gamma_{t+1}=out(TransformerLayer(y_{t+1}))
\end{equation}

\subsubsection{Prediction Module}

During prediction, we use a linear weighted combination of the three scores to obtain the final prediction result:

\begin{equation}
    \hat{r}_{t+1}=output(\alpha_t,\beta_t,\gamma_{t+1})
\end{equation}

\section{Experimental Results and Analysis}
\label{section4}

This section evaluates TLSQKT on three real-world datasets and analyzes its effectiveness for both knowledge tracing and literacy tracing. We first describe the datasets (ASSIST09, Object-math, and Math-literacy) and the comparison baselines (DKT, AKT, DKVMN, CSKT, DTransformer, and QIKT), then detail the experimental protocols and discussions.

\subsection{Datasets}

\begin{table}[htbp]
\begin{threeparttable}
\caption{Statistical information about the three datasets for experiments.}
\begin{center}
\begin{tabular}{c|c|c|c|c|c}
\hline
Dataset           & Students  & Questions & KCs   & Interactions & Literacy \\ \hline
ASSIST09      & 4217 & 26688 & 123 & 346860& -  \\   
Object-math   & 5224 & 16    & 16  & 83584  & - \\ 
Math-literacy & 5224 & 16    & 16  & 83584  & 6  \\  \hline
\end{tabular}
\begin{tablenotes}
    \item Note:The meaning of null value is that the dataset does not contain this.
\end{tablenotes}
\label{dataset}
\end{center}
\end{threeparttable}
\end{table}

The experiment was conducted on three real-world datasets: the public dataset ASSIST09, and two private datasets named Object-math and Math-literacy. All datasets contain student practice sequences along with relevant contextual information. For the public dataset ASSIST09, the data consists of student interactions with the skill builder problem set collected from the Assistment online tutoring platform. This dataset is widely used in knowledge tracing research due to its high applicability. We use the problem-id, skill-id, and correct as the core components of the practice sequences. The Object-math dataset is a private dataset collected from Grade 11 monthly mathematics exams at Changshui High School. We use question-id, knowledge-component-id, and response as the main elements in the sequence data. The Math-literacy dataset is an extended version of Object-math, with an added field literacy-id to indicate the core literacy dimension associated with each question or knowledge component. Table \ref{dataset} presents the statistical summary of these datasets.

\subsection{Baselines}

In the tracing task, we compare our proposed model TLSQKT with several established knowledge tracing models as baseline models to validate its effectiveness:

\begin{itemize}
    \item DKT\cite{DKT}: The first model to introduce deep learning into KT. It uses an LSTM to encode a student’s knowledge state and predict their future responses.
    \item AKT\cite{AKT}: It employs a monotonic attention mechanism to model the distance between the current question and past interactions. It also incorporates the Rasch model to encode these interactions.
    \item DKVMN\cite{DKVMN}: It enhances DKT by using a Dynamic Key-Value Memory Network. It includes a static key matrix to store relationships between questions and knowledge components, and a dynamic value matrix to represent the student’s evolving knowledge state.
    \item CSKT\cite{cskt}: It introduces curriculum-shifted kernel attention to guide learning from short sequences while ensuring accurate prediction for long sequences. It also employs cone-shaped attention to better capture hierarchical relationships among knowledge components.
    \item DTransformer\cite{Dtransformer}: A typical application of the Transformer architecture in KT. It explicitly models students’ mastery of knowledge components and applies a cross-attention mechanism to fuse question and answer interactions for fine-grained knowledge tracing.
    \item QIKT\cite{QIKT}: It utilizes a dual-channel modeling approach that captures learning interactions at both the question level and the knowledge concept level. It predicts students' mastery probabilities for all questions and concepts, offering a more comprehensive diagnostic view.
\end{itemize}

\subsection{Experiment Setup}

During the experiments, the maximum input sequence length was set to 200. Considering the shorter sequences in the private datasets, this maximum length was adjusted to 20 for those cases. We used 80\% of the student data for training and validation, and the remaining 20\% for testing. The models were evaluated using two standard metrics: AUC (Area Under the ROC Curve) and ACC (Accuracy). Early stopping was applied when the AUC score did not improve for 10 consecutive epochs.
All models were implemented using the PyTorch framework and trained on a NVIDIA GTX 1650 Ti GPU.

\subsection{Overall performance}

The results of all models on the two knowledge tracing datasets are presented in Table \ref{KnowledgeTracing}. As shown, our proposed model outperforms all baseline models on the public dataset, demonstrating the effectiveness of our model`s improvements. The integration of the Transformer module allows the model to better capture dependencies within student learning sequences, leading to enhanced prediction performance. However, it is worth noting that the model’s performance slightly decreases on the private dataset. This suggests that the model still faces challenges when dealing with shorter sequences, and is better suited for longer sequential data.

\begin{table}[htbp]
\caption{The experimental results of KT models on the two knowledge tracing datasets.}
\begin{center}
\begin{tabular}{c|c|c|c}
\hline
Model        & Datasets     & AUC    & ACC    \\ \hline
\multirow{2}{*}{DKT \cite{DKT}}            & ASSIST09    & 0.7268 & 0.7512 \\ 
             & Object-math & 0.8504 & 0.7825 \\ \hline
\multirow{2}{*}{AKT \cite{AKT}}            & ASSIST09    & 0.8842 & 0.8209 \\ 
             & Object-math & 0.8456 & 0.7691 \\ \hline
\multirow{2}{*}{DKVMN \cite{DKVMN}}          & ASSIST09    & 0.8796 & 0.8166 \\ 
             & Object-math & 0.8598 & 0.7885 \\ \hline
\multirow{2}{*}{CSKT \cite{cskt}}           & ASSIST09    & 0.8779 & 0.8195 \\ 
             & Object-math & 0.8466 & 0.7752 \\ \hline
\multirow{2}{*}{Dtransformer \cite{Dtransformer}}   & ASSIST09    & 0.7775 & 0.7402 \\ 
             & Object-math & 0.8370 & 0.8040 \\ \hline
\multirow{2}{*}{QIKT \cite{QIKT}}           & ASSIST09    & 0.8967 & 0.8286 \\ 
             & Object-math & 0.8506 & 0.7786 \\ \hline
\multirow{2}{*}{TLSQKT}         & ASSIST09    & 0.9011 & 0.8303 \\ 
             & Object-math & 0.8496 & 0.7763 \\ \hline
\end{tabular}
\label{KnowledgeTracing}
\end{center}
\end{table}

The performance of all models on the literacy tracing dataset is shown in Table \ref{LiteracyTracing}. As illustrated, our model achieves the highest AUC score, demonstrating strong performance in the competency tracing task. This result confirms that our model is capable of effectively transferring to literacy tracing and adapting to this more complex evaluation scenario.

\begin{table}[htbp]
\caption{The experimental results of LT models on the literacy tracing dataset.}
\begin{center}
\begin{tabular}{c|c|c}
\hline
Model        & AUC    & ACC    \\ \hline
DKT \cite{DKT}        & 0.8297 & 0.7632 \\ 
AKT \cite{AKT}        & 0.8124 & 0.7709 \\ 
DKVMN \cite{DKVMN}       & 0.8171 & 0.7664 \\ 
CSKT \cite{cskt}        & 0.8273 & 0.7231 \\ 
Dtransformer \cite{Dtransformer} & 0.8441 & 0.8057 \\ 
QIKT \cite{QIKT}         & 0.8490 & 0.7746 \\ 
TLSQKT          & 0.8515 & 0.7805 \\ \hline
\end{tabular}
\label{LiteracyTracing}
\end{center}
\end{table}

\subsection{Ablation Study}

\begin{figure}[htbp]
	\centering
	\includegraphics[width=1\linewidth]{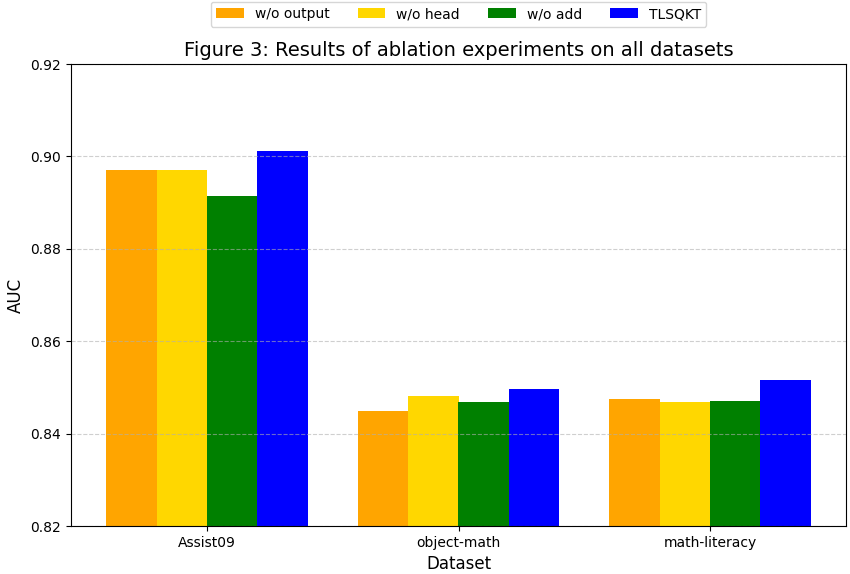}
	\caption{Results of ablation experiments on all three datasets.}
	\label{fig-ablation}
\end{figure}

To verify the necessity of the components in the TLSQKT model, particularly the contribution of the Transformer module, we conducted an ablation study by designing several variants of TLSQKT. "w/o output": This variant removes the Transformer module and feeds the LSTM output directly into a final output layer. The results indicate that using only a simple output layer fails to capture the dependencies between elements in the input sequence. "w/o head": This version replaces the multi-head attention mechanism with a single-head attention to evaluate the impact of multi-head attention. "w/o add": In this setting, the Transformer module is added on top of the existing MLP layer, keeping both components, to assess their combined effect. The complete TLSQKT model is used as the baseline for comparison, and AUC is used as the evaluation metric.
The results of the ablation experiment are presented in Fig. \ref{fig-ablation}. The experiment demonstrate that our model possesses the most appropriate architecture, enabling it to perform more effectively in the competency tracing task.

\section{Conclusions}
\label{section5}

This research proposes a research framework that extends from knowledge tracing to literacy tracing, innovatively transferring knowledge tracing models to the domain of literacy assessment. To better address the challenges of literacy tracing, we introduce a Transformer-based Knowledge Tracing Model for Learning Sequence with Question-Aware Interaction, which builds upon existing models by integrating dual-channel modeling and self-attention mechanisms. We conduct extensive experiments on three real-world datasets, including both public and private datasets. The results demonstrate that TLSQKT achieves superior performance in both knowledge tracing and literacy tracing tasks, confirming the effectiveness and generalizability of the proposed approach.
This research represents a preliminary attempt to transfer knowledge tracing models to the literacy tracing task. While the initial experiments explored the model's performance in this new task, we did not conduct in-depth structural adjustments or develop a detailed adaptation plan. Moreover, a comprehensive literacy assessment framework has yet to be established. Future work will focus on further refining the model architecture and conducting in-depth research on literacy evaluation systems, aiming to build a more robust and systematic approach to literacy tracing.


\bibliographystyle{IEEEtran}
\bibliography{ref,My}

\end{document}